\documentclass[twocolumn,10pt]{article}

\providecommand\apj{ApJ}                 
\providecommand\apjs{Astroph. J. Supp.}                 
\providecommand\aap{Astron. \& Astr.}            
\providecommand\jetpL{JETP Letters}
\providecommand\PRL{Phys. Rev. Lett.}
\providecommand\mnras{MNRAS}
\providecommand\nat{Nature}
\providecommand\prd{Phys. Rev.~D}
\providecommand\cqg{Class. Quant. Grav.}

\providecommand\cqg{CQG}

\providecommand\jetpL{Journal of Experimental and Theoretical Physics Letters}
\providecommand\PRL{Physical Review Letters}

\providecommand\hGpc{\mbox{$h^{-1}$ Gpc}}

\newcommand\rinj{r_{\mathrm{inj}}}  

\usepackage{hyperref}
\usepackage{amsfonts}
\providecommand{\eprint}[1]{\href{http://arxiv.org/abs/#1}{arXiv:#1}}

\begin{document}

\title{Which FLRW comoving 3-manifold is preferred observationally and theoretically?}

\author{BOUDEWIJN F. ROUKEMA}

\date{Toru\'n Centre for Astronomy, 
Nicolaus Copernicus University\\
ul. Gagarina 11, 87-100 Toru\'n, Poland\\
{{\em \small Twelfth Marcel Grossmann Meeting on General Relativity, ed. Thibault Damour, Robert T Jantzen, Remo Ruffini, Singapore: World Scientific}}}

\maketitle

\begin{abstract}
  The lack of structure greater than 10{\hGpc} in Wilkinson Microwave
  Anisotropy Probe (WMAP) observations of the cosmic microwave
  background (CMB) favours compact
  Friedmann-Lema\^{\i}tre-Robertson-Walker (FLRW) models of the
  Universe. The present best candidates based on observations are the
  Poincar\'e dodecahedral space $S^3/I^*$ and the 3-torus $T^3$. 
  The residual gravity effect favours the Poincar\'e space, while
  a measure space argument where the density parameter is a derived
  parameter favours flat spaces almost surely.
\end{abstract}

\newcommand\keywords{observational cosmology; theoretical cosmology; early universe}


\section{FLRW models: curvature and topology}\label{s-flrw}

Comoving space in the Friedmann-Lema\^{\i}tre-Robertson-Walker
(FLRW) models is a 3-manifold 
$M = \widetilde{M}/\Gamma$ \cite{Fried23,Lemaitre31ell,Rob35}.
This can be thought of (i) embedded in a 
higher dimensional euclidean space, e.g. $T^2$ as the surface of 
a doughnut-with-a-hole in $\mathbb{R}^3$; (ii) as a
fundamental domain with identified faces, e.g. a square with 
identified edges for $T^2$ (like in some video games); 
or (iii) as a tiling of the covering space (apparent space)
$\widetilde{M}$ by many copies of the fundamental domain, e.g. 
squares tiling $\mathbb{R}^2$. The group of holonomy transformations
$\Gamma$ identifies multiple images of any single physical object
in $\widetilde{M}$, e.g. $\Gamma = \mathbb{Z}^2$ (linear span of two
vectors) for $T^2$.

\section{Observations}

A first-order argument following immediately from thinking of the
fundamental domain of the 3-manifold is that {\em no physical object
  larger than the fundamental domain can exist}. Hence, observational
statistics representing 
structure (density perturbations) in the Universe should show a
lack of structure on scales larger than the size of the fundamental
domain. In the apparent space, more detailed 
calculations \cite{Star93,Stevens93} show
that an approximate cutoff in structure in the cosmic microwave
background (CMB) should occur. 
This cutoff was suspected in COBE data and was confirmed 
at scales above $\sim 10{\hGpc}$ in 
Wilkinson Microwave Anisotropy Probe (WMAP) maps of the
CMB \cite{WMAPSpergel,Copi07,Copi09}.

Which specific 3-manifold best fits the observations?
Well-proportioned spherical spaces \cite{WeeksWellProp04}, i.e. those
with equal sizes in different fundamental directions, are expected to
more easily fit the WMAP data than other spaces \cite{WeeksWellProp04}.
Among these, the Poincar\'e dodecahedral space, $S^3/I^*$, has
become a particularly good (though disputed) candidate given
the WMAP CMB data \cite{LumNat03,RLCMB04,Aurich2005a,Aurich2005b,Gundermann2005,KeyCSS06,NJ07,Caillerie07,LR08,RBSG08,RBG08}.

On the other hand, while some identified circle \cite{Corn96} analyses
failed to find evidence \cite{CSSK03,KeyCSS06} for the simplest flat
spaces (e.g. $T^3$; infinite space is not simple), several other
analyses find that the $T^3$ model generally provides a better fit to
the WMAP data than infinite flat
models \cite{Aurich07align,Aurich08a,Aurich08b,Aurich09a}.  

\section{Theoretical arguments}

Possible elements of a theory of cosmic topology
include quantum gravity work investigating the
decay from pure quantum to mixed states \cite{Hawking84a}, smooth
topology evolution \cite{DowS98} and some approaches to
deciding which 3-manifold should be
favoured by quantum cosmology \cite{Masafumi96,CarlipSurya04}.

At a much simpler level, a recent heuristic result concerns a
dynamical feedback effect of cosmic topology.  In the presence of a
density perturbation (massive point particle above a homogeneous
background), a residual weak-limit gravitational acceleration effect
exists \cite{RBBSJ06}. Well-proportioned spaces \cite{WeeksWellProp04}
are also ``well balanced'' according to this effect, in the sense that
the first-order effect cancels and the remaining effect is only third
order in the fractional displacement of a test particle from the
massive particle \cite{RBBSJ06,RR09}. What is even more surprising is
that the space that has raised considerable interest in empirical
analyses, the Poincar\'e dodecahedral space $S^3/I^*$, is even
``better balanced'' than the other well-proportioned, well-balanced
spaces. The first and third order terms {\em both} cancel, leaving an
effect dominated by the fifth-order term \cite{RR09}. Thus, the
observational analyses favouring the Poincar\'e space are matched by
this theoretical argument showing that the Poincar\'e space is an
optimal space in terms of the residual gravity effect.  Nevertheless,
although this is an exciting coincidence, it is still a long way from
constituting a physical theory.

In fact, a theoretical argument exists in favour of flat compact
models, by assuming that the density parameter $\Omega$ is a {\em
  derived} rather than fundamental parameter \cite{RB10}. Suppose that
the processes at the exit of the quantum epoch that select a spatial
3-manifold result in a global mass-energy and a Hubble parameter in a
way that is independent of curvature and topology. Then, contrary to
the usual assumption that $\Omega$ is a free parameter, any 3-manifold
(of negative, zero, or positive constant curvature) allows just one value
of $\Omega$ [Eqs~(6), (7), (8), Ref.~\cite{RB10}].
 If, moreover, the injectivity radius $\rinj$
is used to to define a
probability space over the 
set $F$ of compact, comoving, 3-spatial sections of FLRW models,
then the natural measure is the Lebesgue measure, and it is
normalisable, resulting in a probability space.
In this case, flat models should occur with probability one,
i.e. almost surely (a.s.), and non-flat models should occur
with probability zero, i.e. they will a.s. not occur \cite{RB10}. 
This argument is related to the rigidity of curved spaces.

\section{Conclusion}

Both the Poincar\'e dodecahedral space $S^3/I^*$ and the
3-torus $T^3$ are observationally viable candidates for the
spatial section of the Universe.
The residual gravity effect favours the former, while
a measure space argument where the density parameter is a derived
parameter favours compact flat spaces almost surely.

\end{document}